\documentclass[aps,preprint,groupedaddress,showpacs]{revtex4-1}
\usepackage[utf8]{inputenc}
\usepackage[T1]{fontenc}
\usepackage{hyperref}
\usepackage{mathrsfs,amssymb,amstext,amsmath,amsthm,eucal}
\usepackage{mathtools} 
\usepackage[shortlabels]{enumitem}
\usepackage{geometry}
\usepackage{xfrac}
\newcommand{\liealg}[1]{\mathfrak{#1}}
\newcommand{\grad}{\mathrm{d}}
\newcommand{\CC}{\mathbb{C}}
\newcommand{\RR}{\mathbb{R}}

\newcommand{\SO}{\mathrm{SO}}
\newcommand{\so}{\mathfrak{so}}
\newcommand{\SU}{\mathrm{SU}}
\newcommand{\su}{\mathfrak{su}}
\newcommand{\Spin}{\mathrm{Spin}}
\newcommand{\spin}{\mathfrak{spin}}
\newcommand{\dvol}{\mathrm{dvol}} 
\newcommand{\Pz}{P_z}
\newcommand{\Pbz}{P_{\bar{z}}}
\renewcommand{\Re}{\operatorname{Re}}
\renewcommand{\Im}{\operatorname{Im}}
\newcommand{\angles}[1]{\left\langle #1 \right\rangle}

\newcommand{\Cl}{\mathrm{Cl}}
\newcommand{\sect}[1]{\Gamma\left( #1 \right)}

\begin{document}

\author{George Moutsopoulos}
\affiliation{Department of Mathematics, Bogazici University}
\email{gmoutso@gmail.com}
\date{\today}
\title[$\sfrac{1}{4}-$BPS solutions in 3d $\mathcal{N}=16$ and Liouville]{Quarter-BPS solutions in three-dimensional $\mathcal{N}=16$ supergravity and the Liouville equation}

\begin{abstract}
We show how by assuming at least 8 real timelike supersymmetries in the maximally supersymmetric three-dimensional ungauged supergravity and a further simplifying Ansatz, we are naturally led to a pair of Liouville field equations. These are solvable in terms of two meromorphic functions and we present a novel and interesting large class of $\sfrac14$-supersymmetric backgrounds. We also show that there are no solutions that preserve only 6 or 7 real timelike supersymmetries. The solution relies on the classification of complex spinors of Spin(8) to which the problem quickly reduces. 
\end{abstract}
\maketitle
\tableofcontents

\section{Introduction}
The three-dimensional, maximally supersymmetric, ungauged supergravity was found in~\cite{marcus_three-dimensional_1983} and is related to the dimensional reduction of eleven-dimensional supergravity on an internal torus. The size of the  $E_8$ global symmetry and $\mathcal{N}=16$ supersymmetry means that the solution space can be easily probed by assuming those solutions that preserve enough symmetry or supersymmetry. These solutions are interesting for the study of the theory on its own merit as a three-dimensional supersymmetric model, but also  because they oxidize to higher-dimensional supersymmetric string backgrounds. The latter are of special interest in string theory, for instance when the $E_8$-monodromies are not in the automorphisms of the internal torus and so the higher-dimensional geometry is only locally well-defined \cite{Hull:2006va,de_boer_classifying_2014}.

The bosonic content of the three-dimensional maximally supersymmetric supergravity multiplet is a metric and an $E_8/\Spin(16)$ coset scalar. A timelike solution of a supergravity theory is by definition one with a Killing spinor that squares to a timelike vector. Since such Killing spinors are related to the BPS projection equations of supersymmetric branes in the higher-dimensional theory, we will also call them BPS solutions. The timelike solutions of the $\mathcal{N}=16$ theory were classified recently in \cite{de_boer_classifying_2014} by using a novel scheme: each timelike solution is necessarily related to an element $\Pz\in \left( \liealg{e}_8/\spin(16)\right)^{\CC}$ that is nilpotent as an element in $\liealg{e}_8^{\CC}$. There are ten such nilpotent orbits under the action of $\Spin(16)^{\CC}$ that correspond to timelike solutions, which preserve different amounts of supersymmetry, and for each orbit a representative $\Pz$ of the orbit was given that is indeed a solution to the theory.

The classification of \cite{de_boer_classifying_2014} is a novel and ingenious classification scheme, where (a) not all the equations of motion are imposed and the classification becomes algebraic, and (b) the solutions are classified up to the action of $\Spin(16)^{\CC}$ on certain degrees of freedom. That is, one uses the complexification of the local symmetry of the theory that is not a symmetry of the theory and the classification is not on-shell. Finding known solutions that fall into each class is then a separate task, which completes the classification. Our original motivation was an attempt to solve for (all) timelike supersymmetric solutions, where we would use the equations of motion fully and only use the local symmetry of the theory to identify solutions. This is a very difficult problem, if indeed solvable in full generality. One method that follows \cite{de_boer_classifying_2014} is to split each relevant supersymmetric $\Spin(16)^{\CC}$-orbit in  \cite{de_boer_classifying_2014} into its constituent $\Spin(16)$-orbits, each of which is a distinct candidate solution to the theory, but that is a again a daunting task. 

What we present here is a relatively straightforward method, different to the classification of \cite{de_boer_classifying_2014}, in order to address our problem. Namely, in our method we begin by classifying the solutions under the real [\textit{sic}] local symmetry $\Spin(16)$ of the theory - directly. But it is a method with which we soon reach a deadlock and we make two simplifying assumptions in order to proceed. Firstly, we assume enough supersymmetry in order to simplify the problem. More precisely, a quarter supersymmetry reduces some of the scalar  degrees of freedom to spinors of $\Spin(8)$ that we know how to classify. Secondly, a simplifying Ansatz on the remaining scalar degrees of freedom, which is expressed very naturally in our formalism, reduces all equations of motion to two independent two-dimensional Liouville field equations in euclidean signature. A summary of this work is then how the  Liouville equation appears naturally in quarter-BPS solutions of the theory. The Liouville equations are solvable in terms of meromorphic functions and we thus arrive at a large class of 1/4-BPS solutions of 3d maximally supergravity. Our note presents an immediate proposal for future work, which is how to overcome these two assumptions.

Let us give a summary of our derivation. In \S\ref{sec:theory} we begin by presenting the theory and its supersymmetry. In \S\ref{sec:one} we show how the existence of one timelike Killing spinor for a supergravity background, that is one that squares to a timelike vector, implies that the background is ultrastatic in the sense of equation \eqref{eq:ultrastatic1}. The fields and their equations then depend locally only on a two-dimensional spacelike surface. Moreover, the timelike Killing spinor that we assumed is used to construct a projection condition on the scalar degrees of freedom, which is a crucial component to our analysis. In \S\ref{sec:many-timelike-susies} we generalize these results for the existence of at least $n$ timelike Killing spinors. In this case, we derive $n$ compatible projection conditions on the scalar degrees of freedom. For quarter-BPS supersymmetric solutions, that is for 8 out of a possible 32 timelike Killing spinors, the projection conditions are solved explicitly in terms of $\Spin(8)$ spinors. In \S\ref{sec:qbps} we give a large class of quarter-BPS supersymmetric solutions that follows from this analysis. In order to find the solutions, we used the simplifying Ansatz in \eqref{eq:Ansatz} so that the equations reduce to the solvable Liouville equations in \eqref{eq:liouville1}-\eqref{eq:liouville4}. The reduction of quarter-BPS supersymmetric solutions to Liouville equations and the local form of the solution is our main result.
 
In order to arrive at our solution we used the simplifying Ansatz in \S\ref{sec:qbps}. Notwithstanding this, the derivation is extremely straightforward and solvable, but one that has not appeared before. As mentioned in the beginning of this introduction, the large class of solutions in terms of meromorphic functions is then amenable to interesting analyses, in particular with respect to their global properties and the interpretation of the monodromies in the oxidized backgrounds of higher-dimensional supergravity. Had one not needed the Ansatz in the derivation, all timelike $\sfrac14$-solutions would be known. For this reason, the naturalness or generality of this Ansatz is discussed further in section \S \ref{sec:ansatz}. Another point of interest is whether one can derive a similar large class of solutions with less timelike supersymmetry preserved. In section \S\ref{sec:3bps} we show that $\sfrac{3}{16}$-BPS in fact implies $\sfrac{1}{4}$-BPS solutions. We conclude with a short discussion in \S\ref{sec:discussion} and there are two supplementary appendices. The explicit spinor representations of $\Spin(16)$ and $\Spin(8)$ that we use are in terms of exterior algebras, and are first introduced in the second half of section \S\ref{sec:many-timelike-susies}. 

\section{The Theory}
\label{sec:theory}
In this section we introduce the maximally supersymmetric ($\mathcal{N}=16$) three-dimensional supergravity theory \cite{marcus_three-dimensional_1983} and develop the notation that we will use. The global symmetry group of the theory is the split real form of $E_8$. The split real form Lie algebra $e_8$ decomposes symmetrically under the Cartan involution as
\begin{equation}\label{eq:cartan}
\liealg{e}_8 = \spin(16) \oplus \Delta_{16}^+~,
\end{equation}
where $\Delta_{16}^+\cong\RR^{128}$ is the real chiral spinor representation of $\spin(16)$, and $\spin(16)$ is the maximally compact subalgebra of  $\liealg{e}_8$. In order to describe the Lie algebra of  $\liealg{e}_8$, we hence need to give the Lie brackets
\begin{align}
[\spin(16),\spin(16)] &\subseteq \spin(16)~,\\
[\spin(16), \Delta_{16}^+] &\subseteq \Delta_{16}^+~,\label{eq:lie2}\\
[\Delta_{16}^+,\Delta_{16}^+]&\subseteq \spin(16)~.\label{eq:lie3}
\end{align}
The first bracket is the Lie bracket of $\spin(16)\cong\so(16)$, which we may define as the infinitesimal endomorphisms of the real vector space $V_{16}\cong\RR^{16}$ that preserve a positive-definite metric $\eta$ on $V_{16}$. The second bracket is the spin representation $\rho_{\spin}$ of $\spin(16)$ acting on spinors in $\Delta_{16}^+$. Let us assume a representation of the Clifford algebra 
$\Cl(V_{16})$ on $V_{16}$,
\begin{equation}
v\cdot v =-\eta(v,v)\quad\text{for } v\in  V_{16}~. \label{eq:cliffv16}
\end{equation}
For $q\in\spin(16)\cong\Lambda^2 V_{16}$ and $p\in\Delta_{16}^+$, the second Lie bracket is related to the Clifford action $q\cdot p$ as
\begin{equation}\label{eq:spinrep}
[q,p]= \mathrm{\rho}_{\spin}(q) \cdot p = -\frac{1}{4} q \cdot p~.
\end{equation}
The third Lie bracket will be defined in terms of a real, symmetric, positive-definite spin-invariant inner product $\angles{-,-}$ on $\Delta_{16}^+$. For any $p,p'\in\Delta_{16}^+$ and $q\in \spin(16)$ it is symmetric
\begin{align}
\angles{p,p'}&=\angles{p',p}
~,\\\intertext{and spin-invariant}
\angles{q \cdot p, p'}&=- \angles{ p,q \cdot p'}
~.
\end{align}
We then define the third Lie bracket \eqref{eq:lie3} implicitly by
\begin{equation}\label{eq:e8bracket1}
\eta\left([p,p'],q\right)=2\angles{p,q\cdot p'}~.
\end{equation}
Here, $\eta$ was naturally extended to a metric on $\spin(16)\cong \Lambda^2V_{16} \subset V_{16}\otimes V_{16}$. Note that rescaling $\angles{-,-}$ or equivalently rescaling the right-hand side of \eqref{eq:e8bracket1} gives an isomorphic Lie algebra, 
 but a change of sign in \eqref{eq:e8bracket1} 
 defines the maximally compact real form. 
 More generally, we will use the symbols $\spin(d)$, $V_d$,  $\Delta_d$, $\Delta^+_d$ and $\Delta^-_d$ for the $d$-dimensional spin algebra and its vector, spinor, chiral and anti-chiral representations, respectively, for various dimensions $d$. 

The bosonic sector of maximal supergravity in three dimensions is given by a lorentzian mostly-minus metric $g$ on a spin manifold $M$ and a coset map $\mathcal{V}:M\rightarrow E_8/\Spin(16)$. The pull-pack of the bundle $E_8\rightarrow E_8/\Spin(16)$ by the map $\mathcal{V}$ defines a  principal $\Spin(16)$ bundle over $M$ with which we associate the spacetime fermions. The gravitino is a real Rarita-Schwinger field with values in $V_{16}$, and the dilatino is a real spacetime-chiral spinor with values in $\Delta_{16}^-$. The supersymmetries of the theory are given by real spacetime-chiral spinor fields with values in $V_{16}$. Since the spinor representation of $\spin(1,2)\cong\liealg{sl}(2,\RR)$ is two-dimensional, there are in total $2\times16=32$ real local supersymmetries in the theory. 

The action of the full theory and the supersymmetry variations were found in \cite{marcus_three-dimensional_1983}. In this note we are only interested in bosonic backgrounds given by $(g,\mathcal{V})$. In practice, we denote by $\mathcal{V}$ a representative in the coset $E_8/\Spin(16)$, which depends on a choice of a local section of $E_8\rightarrow E_8/\Spin(16)$. That is, we use the so-called gauge formulation of the non-linear sigma model instead of coordinates on $E_8/\Spin(16)$. We decompose the pull-back of the Maurer-Cartan form $\mathcal{V}^{-1}\grad \mathcal{V}$ under \eqref{eq:cartan} and define
\begin{equation}
 \mathcal{V}^{-1}\grad \mathcal{V} = Q+P \in  \, \sect{ \spin(16)\otimes T^{*}M \oplus \Delta_{16}^+\otimes T^{*}M }~.
\end{equation}
Here and in the following, $\sect{E}$ denotes the space of sections of a bundle $E$ over $M$, or of a bundle over $M$ with fiber $E$. The equations of motion are given by the non-linear sigma model action coupled to Einstein gravity without cosmological constant
\begin{equation}\label{eq:action}
S[g,V] = \int \left(- R_g \, \dvol_g  +\angles{ P , \wedge \ast P } \right)~.
\end{equation}
Under a global $\gamma\in E_8$ transformation with local compensating gauge tranformation $h\in \Spin(16)$, that is $\mathcal{V}(x) \mapsto \gamma \mathcal{V}(x) h$, the element $P$ transforms as a spinor $P\mapsto e^{[h,-]}P$ and since $\angles{-,-}$ is $\Spin(16)$-invariant, the action is $E_8$-invariant. 

Let us introduce local spacetime coordinate indices $\mu,\nu=0,1,2$. The Einstein equation and scalars' equation are respectively
\begin{align}
R_{\mu\nu}- \angles{P_{\mu},P_{\nu}}&=0~, \label{eq:Einstein1}\\
\left(\nabla_{\mu}+Q_{\mu}\right)P^{\mu}&=0~, \label{eq:dilaton}
\end{align}
whereas the integrability of $P+Q=\mathcal{V}^{-1}\grad \mathcal{V}$ are
\begin{align}
\grad P + Q \wedge P &= 0 \label{eq:integrability1}~,\\
\grad Q + \frac12 [Q ,Q] &= -\frac12 [P ,P]~.\label{eq:integrability2}
\end{align}
In \eqref{eq:dilaton} and \eqref{eq:integrability1} the connection $Q$ acts in the spin representation $\Delta_{16}^{+}$ of $\spin(16)$ on $P$ and the brackets in the last equation are those of $\liealg{e}_8$, e.g. $[P,P]=[P_{\mu},P_{\nu}]\grad x^{\mu} \wedge \grad x^{\nu}$. Rather than solve for $\mathcal{V}$, we may solve for $P$ and $Q$ subject to the integrability equations \eqref{eq:integrability1}-\eqref{eq:integrability2}. 

Let $\Delta_{(1,2)} \cong \RR^2$ be the Majoranna-Weyl spinor representation of $\Spin(1,2)\cong\mathrm{SL}(2,\RR)$. A Killing spinor
\begin{equation}
\epsilon \in  \sect{\Delta_{(1,2)}\otimes V_{16} }
\end{equation} is a supersymmetry that leaves the gravitino and dilatino, which we have put to zero, invariant. These conditions are respectively
\begin{align}
(\nabla+Q) \epsilon & = 0~, \label{eq:killing1}\\
\left.  P \otimes \epsilon \right|_{\sect{\Delta_{16}^-\otimes \Delta_{1,2}}} & = 0~. \label{eq:dilatino1}
\end{align}
In the first equation, $Q$ acts on $\epsilon$ in the vector representation $V_{16}$ of $\spin(16)$. In the second equation, we use Clifford multiplication in the appropriate slots 
of
\begin{equation}
P \otimes \epsilon\in 
\sect{
\left(T^{*}M \otimes \Delta_{16}^+ \right)\otimes \left(V_{16} \otimes \Delta_{1,3}\right)
} 
\rightarrow 
\sect{
 \Delta_{16}^-\otimes \Delta_{1,2}
} 
~.
\end{equation}
Our conventions for the Clifford algebra $\Cl(TM)$ on $TM$ is
\begin{equation}
X\cdot X  =-g(X,X) \quad\text{for }X \in TM ~.
\end{equation}
Since the Killing spinor $\epsilon$ appears linearly in its defining equations, we will treat it as commuting (Grassmann-even). We will call a $\sfrac{n}{32}$-supersymmetric solution one that preserves $n$ real supersymmetries, i.e. admits $n$ independent Killing spinors, and a quarter-supersymmetric solution is one that admits 8 independent Killing spinors.

Let us comment on the choice of notation used. In a more familiar notation, a vector $v\in V_{16}$ would have 16 components $v^I$, a spinor $p\in\Delta_{16}^+$ has 128 real components $p^A$, $\Cl(V_{16})$ gamma matrices have off-diagonal components $\Gamma^I_{A\dot{A}}$, the spinor inner product is $\angles{-,-}_{AB}=\delta_{AB}$, etc. For instance, the dilatino variation \eqref{eq:dilatino1} would be written in a form such as $\Gamma^I_{A\dot{A}}P_{\mu}^A\gamma^{\mu}\epsilon^I=0$. In order to perform explicit calculations, we will introduce in \S\ref{sec:many-timelike-susies} an alternative representation for $\Cl(V_{16})$, $\Delta_{16}^+$ and $\angles{-,-}$, where the notation of this and the next section are more useful.

\section{One Timelike Supersymmetry}
\label{sec:one}
Assume now a solution that possesses a Killing spinor $\epsilon$. It follows from the closure of the local supersymmetry algebra of a supergravity theory that a Killing vector may always be formed from a Killing spinor. For the theory at hand
\begin{equation}\label{eq:killingv}
K = \eta(\bar\epsilon,\grad x^{\mu} \cdot \epsilon ) \, \partial_{\mu} \in \sect{TM}~,
\end{equation}
is a Killing vector. In the definition of $K$ above, $\bar\epsilon$ is the $\Spin(1,2)=\mathrm{SL}(2,\RR)$ dual of $\epsilon$, $\grad x^{\mu}$ acts by Clifford multiplication of $\Cl(TM)$ on $\epsilon$, and $\eta$ contracts the $V_{16}$ components of the two appearances of the $\epsilon$. The Killing vector $K$ is either null or timelike with respect to the metric $g$. In the first case, the spacetime takes the form of a supersymmetric pp-wave and has been solved completely, see e.g. \cite{de_boer_classifying_2014}. Our focus here is on the timelike case, $g(K,K)>0$. 

The timelike case is also very restrictive in the theory at hand. Indeed, notice that since $\epsilon$ is parallel with respect to $\nabla+Q$ and $Q$ is in $\spin(16)$, then $K$ which is a $\spin(16)$-scalar is also parallel with respect to $\nabla$, i.e. $\nabla K =0$. A $\nabla$-parallel timelike vector $K$ implies that we may bring the metric to the ultrastatic form
\begin{equation}\label{eq:ultrastatic1}
g = \grad t^2 - e^{2\sigma(z,\bar{z})}\grad z \grad \bar{z}~,
\end{equation}
where $K=\partial_t$ is defined everywhere, $t$ is a global function, and the complex coordinate 
\begin{equation}\label{eq:complexcoord}
z = x+i\, y
\end{equation} and $\sigma$ are defined at least locally. The Killing vector $K$ also leaves $\mathcal{V}$ invariant up to a local $\spin(16)$ transformation and we choose a gauge for which $Q_t=P_t=0$ and the
\begin{align}
\Pz&=\frac12 (P_x-i\, P_y) \in \sect{\left( \Delta_{16}^+\right)^{\CC} }~,\\
Q_z&=\frac12 (Q_x-i\, Q_y) \in \sect{\spin(16)^{\CC}}~
\end{align} 
are independent of $t$. 

As was shown in \cite{de_boer_classifying_2014}, the equations of motion and Killing spinor equations simplify for a timelike solution in the complex coordinates \eqref{eq:complexcoord}. The dilatino variation \eqref{eq:dilatino1} is central in the simplification. 
We choose the spacetime gamma matrices
\begin{equation}\label{eq:cl12gamma}
\gamma^0 = \begin{pmatrix} 0 & 1 \\ -1 & 0 \end{pmatrix}~,
\gamma^1 = \begin{pmatrix} 0 & 1 \\ 1 & 0 \end{pmatrix}~\text{ and }
\gamma^2 = \begin{pmatrix} 1 & 0 \\ 0 & -1 \end{pmatrix}~,
\end{equation}
and from the $\spin(1,2)$ spinor components $\epsilon_{\alpha}$, $\alpha=1,2$, we define 
\begin{equation}
\epsilon_z = \epsilon_1 + i\, \epsilon_2 \in \sect{\left( V_{16} \right)^{\CC}}~.
\end{equation}
The dilatino Killing spinor \eqref{eq:dilatino1} then becomes
\begin{equation} \label{eq:dilatino2}
\epsilon_{\bar{z}} \cdot \Pz = 0~,
\end{equation}
where $\epsilon_{\bar{z}} \in \sect{V_{16}^{\CC}}$ acts with Clifford multiplication on $\Pz \in \sect{(\Delta_{16}^+)^{\CC}}$. If we act on \eqref{eq:dilatino2} with $\epsilon_{\bar{z}}$ again, by using the Clifford algebra \eqref{eq:cliffv16} we can show that $\epsilon_{{z}}$ is a complex vector of $\sect{V_{16}^{\CC}}$ that is furthermore null:
\begin{equation}
\epsilon_z\cdot \epsilon_z=0~.
\end{equation}
Complex null vectors can only be of the form
\begin{equation}
\epsilon_z = N (e_1 + i\, e_2)
\end{equation}
where $e_1$ and $e_2$ are real vectors of $V_{16}$ that are orthonormal with respect to $\eta$, 
while $N$ can be chosen to be real and positive. The dilatino Killing spinor equation \eqref{eq:dilatino2} becomes the projection condition
\begin{equation}\label{eq:bps1}
- i\, e_1 \cdot  e_2 \cdot \Pz = \Pz ~.
\end{equation}
Note that here and in the following, what we call a supersymmetry projection condition is a projection on $\Pz$ with the projection operator constructed out of the Killing spinor. This is different to the usual notion of a BPS projection in higher dimensions, that is a projection on the Killing spinor with the projection operator constructed out of a supersymmetric brane worldvolume. In our case, we use the projection conditions to restrict $\Pz$. We may now turn to the equations of motion for a timelike solution.

Let us extend the inner product $\angles{-,-}$ on $\Delta_{16}^{+}$ to the hermitian inner product on  $\left(\Delta_{16}^{+}\right)^{\CC}$ that we also denote by $\angles{-,-}$,
\begin{equation}
\angles{p ,i\, p'} = \angles{-i \, p, p'} = i\, \angles{p , p'} \quad \text{for }p,p'\in \left(\Delta_{16}^{+}\right)^{\CC}~.
\end{equation}
Because of the condition \eqref{eq:bps1} and that $\angles{-,-}$ is $\spin(16)$-invariant, we may show that
\begin{equation}\label{eq:pzp}
\angles{\Pbz,\Pz} = \angles{\Pbz, i \, e_1 \wedge e_2 \cdot \Pz} = \angles{i\, e_1 \wedge e_2 \cdot \Pbz, \Pz} = - \angles{\Pbz,\Pz} = 0~.
\end{equation} 
The Ricci tensor for the ultrastatic metric has $R_{tt}=0$,  $R_{xx}=R_{yy}$ and $R_{xy}=0$ which is consistent with the Einstein equation of motion \eqref{eq:Einstein1} with $P_t=0$ and $\angles{\Pbz,\Pz}=0$ as in \eqref{eq:pzp}. The only remaining non-trivial component of the Einstein equation of motion is
\begin{equation}
\label{eq:Einstein2}
- 2 \partial_z \partial_{\bar{z}} \sigma = \angles{\Pz,\Pz}~.
\end{equation}
The equation of motion \eqref{eq:dilaton} for $\Pz$ and the integrability equation \eqref{eq:integrability1} for $\Pz$ are respectively the real and imaginary part of the single complex equation
\begin{equation}\label{eq:DP1}
\left( \partial_{\bar{z}} + Q_{\bar{z}} \right) \Pz = 0~.
\end{equation}
Finally, the integrability equation \eqref{eq:integrability2} for $Q$ is
\begin{equation} \label{eq:integrability3}
\Im \left( \partial_{\bar{z}} Q_z+ \frac12 [Q_{\bar{z}},Q_z] \right) = 
\frac12 i [\Pbz,\Pz]
~.
\end{equation}
Here and in the following, we have extended complex-bilinearly the bracket $[-,-]$ on  $\liealg{e}_8^{\CC}$.

The $t$-component of the gravitino Killing spinor equation  \eqref{eq:killing1} is simply that the Killing spinor is $t$-independent. Interestingly, the integrability of the gravitino Killing spinor equation in the $x-y$ components is automatically satisfied provided the dilatino Killing spinor equation \eqref{eq:dilatino2}, the Einstein equation \eqref{eq:Einstein2} and the integrability equation \eqref{eq:integrability3} for $Q$ hold, see \cite{de_boer_classifying_2014} for the proof. A timelike solution of the theory is thus subject to the equations of motion \eqref{eq:Einstein2}, \eqref{eq:DP1}, \eqref{eq:integrability3} and the projection condition \eqref{eq:bps1}. We note that \eqref{eq:Einstein2} is the only one that involves the conformal factor $e^{2\sigma}$ of the ultrastatic metric and we may thus focus on the rest.

In the next section we will also introduce explicit spinor representations, and it will be useful to denote complex conjugation on $\Pz$ as an anti-linear involution $C$, i.e.
\begin{equation}\label{eq:PzbarCPz}
\Pbz = C( \Pz )~.
\end{equation} 
Let us extend complex-bilinearly the metric $\eta$ on $\left(V_{16}\right)^{\CC}$ and the Lie bracket $[-,-]$ on $\liealg{e}_8^{\CC}$ in both slots, as we did in \eqref{eq:integrability3}. The Lie bracket of $\liealg{e}_8^{\CC}$ on $\Lambda^2 \left( \Delta_{16}^+\right)^{\CC} \rightarrow \spin(16)^{\CC}$ as defined in \eqref{eq:e8bracket1} can be rewritten as
\begin{equation}\label{eq:e8bracket2}
\eta\left([p,p'],q\right)=2\angles{C(p),q\cdot p'}~,\end{equation}
for any $p,p'\in\left(\Delta_{16}^+\right)^{\CC}$ and $q\in\spin(16)^{\CC}$. We will use this formula to calculate the Lie bracket in \eqref{eq:integrability3}.

\section{Many Timelike Supersymmetries}
\label{sec:many-timelike-susies}
In this section, we investigate the consequences of the existence of more than one Killing spinor, generalizing the projection \eqref{eq:bps1} on $\Pz$ for one Killing spinor. We also introduce an explicit $\Delta_{16}^+$ representation in terms of an auxiliary vector space $U$, in which we write the conditions on $\Pz$. We observe that a timelike supersymmetric solution preserves an even amount of real supersymmetry, because if $\epsilon_z$ is a Killing spinor then so is $i\,\epsilon_z$. This is clear from the dilatino Killing spinor equation \eqref{eq:dilatino2}, which defines timelike Killing spinors subject to the equations of motion  \eqref{eq:Einstein2}-\eqref{eq:integrability3}. We henceforth assume a timelike solution that preserves $2n$ real supersymmetries.

Let us then assume $n$ linearly independent over $\CC$ Killing spinors $\epsilon_z^i\in \sect{V_{16}^{\CC}}$, $i=1,\ldots n$. For each Killing spinor equation, i.e. $\epsilon_{\bar{z}}^i\cdot \Pz=0$, we Clifford multiply with $\epsilon_{\bar{z}}^j$, symmetrize over $(i,j)$ and use the Clifford algebra \eqref{eq:cliffv16}. We may thus show that the $\epsilon_z^i$ are complex, null and orthogonal to each other: 
\begin{equation}\label{eq:allnull}
\eta(\epsilon_z^i,\epsilon_z^j)=0~.
\end{equation}
Let us use an orthonormal basis $e_I$, $I=1,\ldots, 16$, of $V_{16}$. We may use $\SO(16)$ to rotate the $\epsilon_z^i$ to the canonical form
\begin{equation}\label{eq:LD}\begin{aligned}
\epsilon_z^1 &= N_{11}\left( e_1 + i\, e_9 \right) ~,\\
\epsilon_z^2 &= N_{21} \left( e_1 + i\, e_9 \right) + N_{22}\left( e_{2} + i\, e_{10} \right) ~,\\
& \vdots \\
\epsilon_z^n &= N_{n1}  \left( e_1 + i\, e_{9} \right) + N_{n2}\left( e_{2} + i\, e_{10} \right) + \cdots + N_{nn}  \left( e_{n} + i\, e_{n+8} \right)~.
\end{aligned}\end{equation}
Indeed, assuming we have rotated the first $(n'-1)<n$ Killing spinors, the stabilizer of these is $\SO(16-2(n'-1))$ that can be used to rotate the real and imaginary part of the next Killing spinor so that it is in the span of the previous Killing spinors plus $e_{n'}$ and $e_{n'+8}$. In this process, the coefficients in \eqref{eq:LD} are restricted by \eqref{eq:allnull}, and we arrive at a generalization of the left-diagonal decomposition of a matrix. 
This proof by construction is complete provided the diagonal coefficients $N_{ii}$ are not zero, which we show in the appendix \ref{app:lineps}. Since $\Spin(16)$ is a local symmetry of the theory, we are still studying the same timelike solution.

The Killing spinor equations \eqref{eq:dilatino2} for $\epsilon^i_z$ in the canonical form \eqref{eq:LD} are equivalent to the 
compatible supersymmetry projection conditions
\begin{equation}\label{eq:bps2}
-i\, e_{i} \cdot  e_{i+8} \cdot  \Pz = \Pz ~, \quad i=1,2,\ldots ,n~.
\end{equation}
Note that these are projection conditions on $\Pz$, and not projections on the Killing spinor $\epsilon_z$. In writing the projection conditions, we have used $\Spin(16)$, which rotated the Killing spinors and $\Pz$. The remaining freedom in $\Spin(16)$ is the group $U(1)^n \times \Spin(16-2n)$ generated by
\begin{equation}
\left\{ e_i \wedge e_{i+8} \right\}_{1\leq i \leq n}
\oplus 
\left\{ e_{i} \wedge e_{j+8},\, e_{i} \wedge e_{j} ,\, e_{i+8} \wedge e_{j+8}   \right\}_{n<i,j\leq 8}  ~.
\end{equation} 
It leaves the projections in \eqref{eq:bps2} invariant and acts on $\Pz$ as a complex spinor of $\Spin(16-2n)$ with weights under the $U(1)$'s given by \eqref{eq:bps2}.

The main result of this section is \eqref{eq:bps2}, but in order to proceed and solve the projection conditions we need an explicit solution to the projection conditions, and so we need an explicit representation of $\left(\Delta_{16}^+\right)^{\CC}$ to which $\Pz$ belongs. We use a representation of $\mathrm{Cl}(V_{16})$ in terms of $\Lambda^{*}U^{\CC}$, the exterior algebra of the complexification of an 8-dimensional real metric vector space $U$,
\begin{equation}
U = \RR^8\angles{\tilde{e}_1,\ldots, \tilde{e}_8}~,
\end{equation}
where the basis vectors $\tilde{e}_i$, $i=1,2,\ldots,8$, are to be thought of as orthonormal. Let ${p}\in \Lambda^{*}U^{\CC}$ be a complex form. The representation of the Clifford algebra is given by
\begin{subequations}\label{eq:clifford1}\begin{align}
e_i \cdot {p} &= \tilde{e}_i \wedge {p} - i_{\tilde{e}_i}{p} ~,\\
e_{8+i} \cdot {p} &= i\left( \tilde{e}_i \wedge {p}  + i_{\tilde{e}_i}{p}\right)~~,
\end{align}\end{subequations}
for $i=1,\dots,8$. One may indeed confirm that \eqref{eq:cliffv16} is satisfied
\begin{equation}\left(e_I \cdot e_J+ e_J \cdot e_I\right)\cdot {p} = -2 \delta_{IJ} \,  {p} 
\end{equation}
for all $I,J=1,\ldots,16$. The form ${p}$ is a complex spinor of $\Spin(16)$ with indefinite chirality, i.e. $\Lambda^{*}U^{\CC} \cong \Delta_{16}^{\CC}\cong \CC^{256}$.

We denote by $\dvol_8=\tilde{e}_1\wedge \tilde{e}_2 \wedge\cdots \wedge \tilde{e}_8$ the volume form of $U$ and by $\ast_8$  the Hodge star. A hermitian inner product on $\Lambda^{*}U^{\CC}$ is defined by
\begin{equation}
\angles{p,p'} = \left.\left( p^{*} \wedge \ast_8 p' \right)\right|_{\dvol_8}~,
\end{equation}
for any $p,p' \in \Lambda^{*}U^{\CC}$. Note that here we use complex conjugation $p^{*}$ with respect to the real space $U$ and is not the same as the complex conjugation $C$ defined in \eqref{eq:PzbarCPz}. By using the identity $\tilde{e}_i \wedge \ast_8 {p}= \ast_{8} i_{\tilde{e}_i}{p}$ and \eqref{eq:clifford1}, one may show that the $e_I$ are anti-hermitian wih respect to $\angles{-,-}$,
\begin{equation}\label{eq:dirac1}
\angles{e_I \cdot p, p'} = - \angles{p, e_I \cdot p'}~.
\end{equation}
It follows that $\angles{-,-}$ is $\Spin(16)$-invariant,
\begin{equation}
\angles{e_I \wedge e_J\cdot p, p'} = - \angles{p, e_I  \wedge e_J\cdot p'}~,
\end{equation}
where by definition $e_I  \wedge e_J\cdot p=\left(e_I \cdot e_J  -e_J \cdot e_I \right)\cdot p$. An anti-linear involution on $\Lambda^{*}U^{\CC}$ is given by $C$:
\begin{equation}\label{eq:reality1}
C({p}) = \left(\prod_{i=1}^8 e_i \right) {p}^{*} = \ast_8 {p}^{*}~.
\end{equation}
One may show that $e_I \cdot C({p})= -C(e_I \cdot {p})$ and so $C$ commutes with $\Spin(16)$. Real spinors of $\Spin(16)$ are thus forms in $\Lambda^{*}U^{\CC}$ that satisfy the reality condition ${p}=C({p})$. 
The volume form $\dvol_{16}$ in $\mathrm{Cl}(V_{16})$ is represented by
\begin{equation}\label{eq:chirality1}
\frac{1}{16!}\dvol_{16} := \prod_{i=1}^8 e_i \cdot e_{i+8} \cdot =  \prod_{i=1}^8\left( 1-2 \tilde{e}_i \wedge i_{\tilde{e}_i} \right)
\end{equation}
that has positive eigenvalues for even-degree forms ${p}\in\Lambda^{\text{even}}U^{\CC}$. Chiral spinors of $\Spin(16)$ are thus even-degree complex forms, i.e. $\left(\Delta_{16}^+\right)^{\CC} \cong \Lambda^{\text{even}}U^{\CC} \cong \CC^{128}$. 

We may finally return to the projection conditions \eqref{eq:bps2} for
\begin{equation}
\Pz \in \sect{\Lambda^{\text{even}} U^{\CC}}~
\end{equation}
and write them in the explicit representation that we just introduced. 
By using \eqref{eq:clifford1}, they become
\begin{equation}
\left( 1- 2 \tilde{e}_i \wedge i_{\tilde{e}_i} \right) \Pz = -\Pz~, \quad \text{for }i=1,\ldots n,
\end{equation}
or $ \tilde{e}_i \wedge i_{\tilde{e}_i} \Pz = \Pz$ for $i=1,\ldots n$, which implies
\begin{equation}
\label{eq:reduction1}
\Pz = \tilde{e}_1 \wedge \tilde{e}_2 \wedge \cdots \wedge \tilde{e}_n \wedge \Pz^0 ~.
\end{equation}
Then $\Pz^0$ is an even-degree complex form if $n$ is even, or odd-degree complex form if $n$ is odd, with legs in the remaining directions of $U^{\CC}$. Let us consider the case where $n$ is even. By repeating the spinor representation but with $U$ replaced by
\begin{equation}
W = \RR^{8-n}\angles{ \tilde{e}_{n+1} , \cdots, \tilde{e}_8} \subset{U}~,
\end{equation}
and by using equivalent definitions as in  \eqref{eq:clifford1}, \eqref{eq:dirac1}, \eqref{eq:reality1} and \eqref{eq:chirality1},
the unknown degrees of freedom in $\Pz$ are in
\begin{equation}
\Pz^0\in \sect{\Lambda^{\text{even}}W^{\CC}} \cong \sect{\left(\Delta_{16-2n}^+\right)^{\CC}}~,
\end{equation}
that is a complex chiral spinor of $\Spin(16-2n)\subset\Spin(16)$. We may then use the remaining local symmetry $\Spin(16-2n)$ to fix $P_z^0$ to a given form.

\section{Quarter-BPS Solutions}
\label{sec:qbps}
In this section we will focus on $n=4$. 
That is, we assume a solution with at least 4 complex Killing spinors that preserve 8 real supersymmetries out of a possible 32. The group $\Spin(8)$ acts transitively on $\left(\Delta_8^+\right)^{\CC}$ up to two scales. We will use the classification of $\Spin(8)$-orbits of spinors in order to fix ${P}_z^0$ and $\Pz$ in \eqref{eq:reduction1}. However, in order to solve the equations of motion, we will need one further assumption on $Q_z$, which we make towards the end of this section. 

We use the spin representation of $\Spin(8)$ in terms of $\Lambda^{*}W^{\CC}$, where
\begin{equation}
W = \RR^{4}\angles{ \tilde{e}_{5} ,  \tilde{e}_{6} ,  \tilde{e}_{7} , \tilde{e}_8}~.
\end{equation}
It is well-known that a real chiral spinor in $\Delta_8^+$ is in the same $\Spin(8)$-orbit as $1+\tilde{e}_5\wedge \tilde{e}_6\wedge \tilde{e}_7 \wedge \tilde{e}_8$ up to scale with stability $\Spin(7)^+$, under which the representations of $\Spin(8)$ reduce as follows:
\begin{align}
\Spin(8)  &\stackrel{\Spin(7)^+}{=} \Spin(7)^+ \ltimes V_7 ~,\\
\Delta_8^+ &\stackrel{\Spin(7)^+}{=} \RR\angles{1+\tilde{e}_5\wedge \tilde{e}_6\wedge \tilde{e}_7 \wedge \tilde{e}_8} \oplus V_7
~,\label{eq:decompose1}
\\
\Delta_8^{-} &\stackrel{\Spin(7)^+}{=} V_8 \stackrel{\Spin(7)^+}{=} \Delta_7
~.\label{eq:delta7}\end{align}
The last equation can be seen as a consequence of $\Spin(8)$ triality. The vector representation $V_7$ of $\Spin(7)^+$ is spanned by
\begin{equation}
V_7 = \RR\angles{i(1-\tilde{e}_5\wedge \tilde{e}_6\wedge \tilde{e}_7 \wedge \tilde{e}_8)} \oplus \left\{\lambda
 \in \Lambda^2 W^{\CC}: \lambda=\ast_8 \lambda^{*} \right\}~,
\end{equation} 
while $\Delta_7$ in \eqref{eq:delta7} is the real 8-dimensional spin representation of $\Spin(7)^+$. Since we need to fix a complex spinor $P_z^0$, we use the stabilizer $\Spin(7)^+$ to fix a second real spinor in $\Delta_8^+$.

A second spinor in $\Delta_8^+$ decomposes as in \eqref{eq:decompose1} with a component along $1+\tilde{e}_5\wedge \tilde{e}_6\wedge \tilde{e}_7 \wedge \tilde{e}_8$ and a part in $V_7$, but $\Spin(7)^+$ acts on the vector representation $V_7$ transitively. The part in $V_7$ can thus be brought to $i(1-\tilde{e}_5\wedge \tilde{e}_6\wedge \tilde{e}_7 \wedge \tilde{e}_8)$ with stability $\Spin(6)=\SU(4) \subset \Spin(7)^+$. The representations of $\Spin(7)^+$ decompose under $\SU(4)$
\begin{align}
\Spin(7) & \stackrel{\SU(4)}{=} \SU(4) \ltimes \CC^4~,\\
V_7 & \stackrel{\SU(4)}{=} \RR\angles{i(1-\tilde{e}_5\wedge \tilde{e}_6\wedge \tilde{e}_7 \wedge \tilde{e}_8)} \oplus V_6 ~,\\
\Delta_7 & \stackrel{\SU(4)}{=} \CC^4~.
\end{align}
Here $V_6$ is the real 6-dimensional representation of $\SO(6)$ and $\CC^4$ is the chiral spinor representation of $\Spin(6)$. 

It follows from the classification of $\Spin(8)$ spinors that we may fix the real and imaginary parts of $\Pz^0\in \sect{\left(\Delta_8^+\right)^{\CC}}$ under $\Spin(8)$ to be 
\begin{align}
\Re\left(\Pz^0\right) &= \tilde{a} \left( 1+  \tilde{e}_5\wedge \tilde{e}_6\wedge \tilde{e}_7 \wedge \tilde{e}_8 \right) ~,\\
\Im\left(\Pz^0\right) &= \tilde{b}  \left( 1+  \tilde{e}_5\wedge \tilde{e}_6\wedge \tilde{e}_7 \wedge \tilde{e}_8 \right) - i \,\tilde{c}  \left( 1 - \tilde{e}_5\wedge \tilde{e}_6\wedge \tilde{e}_7 \wedge \tilde{e}_8 \right)~,
\end{align}
with $\tilde{a},\tilde{b},\tilde{c}\in\RR$. 
It is convenient to write instead $\Pz^0$ in the more relaxed form
\begin{equation}\label{eq:fixing1}
\Pz^0 = a+b\,\tilde{e}_5\wedge \tilde{e}_6\wedge \tilde{e}_7 \wedge \tilde{e}_8~,
\end{equation}
for generic $a,b\in\CC$. The $U(1)$ subgroup of $\Spin(8)$ generated by
\begin{equation}\label{eq:ua}
L_1 ={\frac12} \sum_{i=5}^8   e_i \wedge e_{i+8}  \stackrel{\rho_{\spin}}{\longmapsto}
i\left( 1 - \frac12 \sum_{i=5}^8  \tilde{e}_i \wedge i_{\tilde{e}_i} \right)~,
\end{equation}
acts on $\Pz$ by sending $(a,b)\mapsto (ia,-ib)$. Here we used the spin representation image in \eqref{eq:spinrep} in terms of the Clifford representation \eqref{eq:clifford1}. By using $e^{t\, L_1}$, we can make either $a$ or $b$ real, which would leave us again with three real parameters. 
The subgroup $\mathrm{SU}(4)\subset \Spin(8)$ that leaves $\Pz^0$ invariant is generated by the Lie algebra elements
\begin{multline} \label{eq:su4}
\frac12 \sum_{i,j=1}^4 A_{ij} \left( e_{i+4} \wedge e_{j+4} + e_{i+12} \wedge e_{j+12} \right)
-
\frac12 \sum_{i,j=1}^{4} B_{ij} \left( e_{i+4} \wedge e_{j+12} + e_{j+4} \wedge e_{i+12} \right)\\
\stackrel{\rho_{\spin}}{\longmapsto}
 \sum_{i,j=1}^4 \left( A_{ij} + i \, B_{ij} \right) \tilde{e}_{i+4} \wedge i_{\tilde{e}_{j+4}}
\end{multline}
where
\begin{equation}
A_{ij}=-A_{ji}~,\quad
B_{ij}=B_{ji}~\text{ and }\quad
\sum_{i=1}^4 B_{ii}=0~,
\end{equation}
and we used the spin representation \eqref{eq:spinrep} and the Clifford representation \eqref{eq:clifford1}, 
see also the appendix \ref{app:spinhelp} for such expressions. If $\Re(\Pz^0)$ is proportional to $\Im(\Pz^0)$, that is if $|a|=|b|$, then the stability subgroup of $\Pz^0$ in $\Spin(8)$ enhances to $\Spin(7)^+$.

It is time to turn to the element 
\begin{equation}\label{eq:fixing2}
\Pz =  \tilde{e}_1 \wedge \tilde{e}_2 \wedge   \tilde{e}_3 \wedge \tilde{e}_4 \wedge \left(  a + b \, \tilde{e}_5\wedge \tilde{e}_6\wedge \tilde{e}_7 \wedge \tilde{e}_8 \right)
\end{equation} and since $Q_z$ belongs to the complexified $\Spin(16)^{\CC}$, it is indeed convenient to allow $a,b\in\CC$. There are two $U(1)$'s in $\Spin(16)$ that are central to our analysis. The first $U(1)$ is generated by 
$L_1$ that was defined in \eqref{eq:ua} and acts on $\Pz$ by sending $(a,b)\mapsto(ia ,- ib)$. The second $U(1)$
is generated by
\begin{equation}\label{eq:ub}
L_2 = \frac12 \sum_{i=1}^{4} e_{i} \wedge e_{i+8} 
 \stackrel{\rho_{\spin}}{\longmapsto}
 i\left( 1 - \frac12 \sum_{i=1}^4 \tilde{e}_i \wedge i_{\tilde{e}_i} \right)~
\end{equation}
and acts on $\Pz$  sending $(a,b)\mapsto(-ia , -ib)$.
We promote the linear analysis to local sections over the manifold and $a,b$ become functions of $(z,\bar{z})$. Let us then write the \textit{Ansatz}
\begin{equation}\label{eq:Ansatz}
Q_z = q_1\, L_1 + q_2\, L_2
\end{equation}
for complex functions $q_1(z,\bar{z})$ and $q_2(z,\bar{z})$. The equation of motion  \eqref{eq:DP1} for $\Pz$ and the integrability of $Q_z$ \eqref{eq:integrability3} via \eqref{eq:e8bracket2} become
\begin{align}\label{eq:liouville1}
&\partial_{\bar{z}} a + i \bar{q}_1 a - i \bar{q}_2 a =0~,\\
&\partial_{\bar{z}} b - i \bar{q}_1 b - i \bar{q}_2 b =0~,\\
&2\Im\left(\partial_{\bar{z}} q_1 \right) = i \angles{\Pz, L_1 \Pz} =  |b|^2-|a|^2~,\\
&2\Im\left(\partial_{\bar{z}} q_2 \right) = i \angles{\Pz, L_2 \Pz} = |b|^2+|a|^2~.\label{eq:liouville4}
\end{align}
We may use the $U(1)\times U(1)\subset \Spin(16)$ symmetry that is generated by $L_1$ and $L_2$ in order to set $a$ and $b$ real.
If both $a$ and $b$ are non-zero, we may substitute the equations for $\partial_{\bar{z}} q_1$ and $\partial_{\bar{z}} q_2$ in order to derive the quadratic equations
\begin{align}
 \partial_z\partial_{\bar{z}} \ln a & =  a^2~,\\
 \partial_z\partial_{\bar{z}} \ln b & =  b^2~.
\end{align}
This is a pair of Liouville's field equation. If one of $a$ or $b$ is zero, we only get one Liouville field equation.

The Ansatz \eqref{eq:Ansatz} is interesting, as far as {test solutions} go, because it leads to a local closed-form solution:
\begin{align}\label{eq:sol1}
a^2 &= \frac{|\partial_zf_1|^2}{\left(1-|f_1|^2\right)^2} = -\partial_z \partial_{\bar{z}} \ln \left(1-|f_1(z)|^2\right) ~,\\
b^2 &=  \frac{|\partial_zf_2|^2}{\left(1-|f_2|^2\right)^2} = -  \partial_z \partial_{\bar{z}} \ln \left(1-|f_2(z)|^2\right) ~,\\
e^{2\sigma}&= \left(1-|f_1(z)|^2\right)^{} \left(1-|f_2(z)|^2\right)^{}~,\label{eq:sol3}
\end{align}
for two arbitrary functions $f_{1,2}(z)$ that are defined locally on some patch. In solving the equations of motion \eqref{eq:Einstein2}-\eqref{eq:integrability3}, we have fixed the conformal Liouville symmetry of the problem completely. 

\section{The Ansatz}
\label{sec:ansatz}
The solutions \eqref{eq:sol1}-\eqref{eq:sol3} are a very large class of local solutions that are interesting to study. For instance, we may generalize the local holomorphic functions to meromorphic functions on an $N$-punctured two-sphere, with $\infty$ representing spatial infinity and the poles representing compatible $\sfrac{1}{4}-$BPS states. One may then study their properties such as coset charges, asymptotics, monodromies, etc. Yet another point of interest, which is relevant to this work, is how general the Ansatz \eqref{eq:Ansatz} is and whether it can be somehow relaxed. We were not able to show that the Ansatz is the most general solution, but at the same time we did not find any straightforward way of generalizing it. In this section, we present  a few details about the naturalness of the Ansatz in our formalism.

Let us not assume \eqref{eq:Ansatz}, but the element $\Pz$ is brought to the form \eqref{eq:fixing2} on any local patch at will. It is convenient to rewrite  the equation of motion \eqref{eq:DP1} for $\Pz$ and the integrability of $Q_z$ \eqref{eq:integrability3}, again here:
\begin{align}
\partial_{\bar{z}} \Pz + Q_{\bar{z}} \Pz &= 0~, \label{eq:dp2}
\\
\Im \left( \partial_{\bar{z}} Q_z+ \frac12 [Q_{\bar{z}},Q_z] \right) &= 
\frac12 i [\Pbz,\Pz]
~.\label{eq:integrability4}
\end{align}
Having fixed $\Pz$ as in \eqref{eq:fixing2}, then due to the terms $\partial_{\bar{z}}\Pz$ and $Q_{\bar{z}}\Pz$ in \eqref{eq:dp2}, the connection one-form $Q_{\bar{z}}$ has to belong to the subalgebra of $\spin(16)^{\CC}$ that preserves the form of $\Pz$ in terms of generic $(a,b)\in\sect{\CC^2}$. That is, $Q_{\bar{z}}$ has to effectively act on $(a,b)$ as a $\CC$-linear endomorphism of $\sect{\CC^2}$, but it may also have components in the stabilizer of $\Pz$ in $\spin(16)^{\CC}$. There is no other possibility. Note that the part of $Q_{\bar{z}}$ that acts effectively on $(a,b)$ can only be defined modulo the stabilizer. A linear algebra calculation shows that the full subalgebra is generated by the $L_1$ and $L_2$ that act effectively on $\Pz$ and the stabilizer of $\Pz$ in $\spin(16)^{\CC}$. That is, modulo the stabilizer the effective action is only given by the action of $L_1$ and $L_2$. The Ansatz \eqref{eq:Ansatz} assumes that $Q_z$ only has components in $L_1$ and $L_2$.

The stability subalgebra of $\Pz$ in $\spin(16)^{\CC}$, in which a more general putative solution for $Q_{\bar{z}}$ might have some components, is generated by three types of elements: 
\begin{enumerate}[(a)]
\item elements in the complexification of $\liealg{u}(1)^3\subset \liealg{u}(1)^4 \subset \liealg{u}(1)^4 \oplus \spin(8)$ that are generated by
\begin{equation}\label{eq:typea}\begin{aligned}
 e_1 \wedge e_9 -  e_2 \wedge e_{10} ~, \quad e_1 \wedge e_9 -  e_3 \wedge e_{11}~\text{ and }\quad e_1 \wedge e_9 -  e_4 \wedge e_{12}~,
\end{aligned}\end{equation}
which anninhilate $\Pz$ because of the projection conditions, 
\item elements in the complexification $\liealg{su}(4)^{\CC}=\liealg{sl}(4,\CC)$, where $\liealg{su}(4)$ was defined in \eqref{eq:su4}, and 
\item  elements generated by
\begin{equation}\label{eq:typec}
\sum_{i,j=1}^4\eta_{ij}  \left( e_{i+4} \wedge e_{j+4} - e_{i+12} \wedge e_{j+12} \right) +  \sum_{i,j=1}^{4} \zeta_{ij} \left( e_{i+4} \wedge e_{j+12} - e_{j+4} \wedge e_{i+12} \right)~,
\end{equation}
provided that $\eta$ and $\zeta$ are related as
\begin{equation}\label{eq:twomore1}
a \, \zeta - b \ast_4 \zeta + i \,a\, \eta + i \,b \ast_4 \eta = 0~
\end{equation}
where we think of $\eta$ and $\zeta$ as complex two-forms in $\Lambda^2 W^{\CC}$.
\end{enumerate}
The elements of type (a) and (b) with real coefficients also describe the real stability subalgebra $\liealg{u}(1)^3\oplus\su(4)$ of $\Pz$ in $\spin(16)$. The elements of type (c) have non-trivial solutions and, in contrast to the stability generators of type (a) and (b), they depend on the $a$ and $b$ in $\Pz$. For instance, if we write the dual and anti-self-dual parts of $\eta$ and $\zeta$ as $\eta^{\pm}$ and $\zeta^{\pm}$ respectively, then \eqref{eq:twomore1} becomes
\begin{equation}\label{eq:twomore2}
i(a\mp b) \zeta^{\pm} = (a \pm b) \eta^{\pm}~.
\end{equation}
There is no purely real solution of \eqref{eq:twomore2}, that is a solution with $\eta^{*}=\eta$ and $\zeta^{*}=\zeta$, unless $|a|=|b|$. Nevertheless, $Q_z$ might have components in $\spin(16)^{\CC}$ and in particular of type (c). 

The real holonomy that is generated by $Q_z$ is in the curvature 
\begin{equation}
R_{xy}^Q = \partial_{x} Q_{y}- \partial_{y} Q_{x} + [Q_x,Q_y]
\end{equation} that is on the left-hand side of \eqref{eq:integrability4}. However, by using the right-hand side of \eqref{eq:integrability4}, we can show that $R_{xy}^Q$ does not have any components in the directions generated by the \textit{real} stabilizer of $\Pz$ in $\spin(16)$. Indeed, for $q\in \spin(16)$ such that $q\cdot \Pz = 0$, we have
\begin{equation}\label{eq:e8bracketz}
\eta(q,[\Pbz,\Pz]) = 2 \angles{ P_{{z}} , q\cdot \Pz} = 0~,
\end{equation}
where the first equality is due to the definition in \eqref{eq:e8bracket2}. 
By using the expressions in the appendix \ref{app:spinhelp} and \eqref{eq:e8bracket2}, we may ultimately show that $R_{xy}^Q$ only has components in $L_1$ and $L_2$:
\begin{equation}\label{eq:RL12}
R_{xy}^Q = \left(|b|^2 - |a|^2 \right) L_1 + \left( |a|^2+|b|^2 \right) L_2~.
\end{equation}
However, the assumption \eqref{eq:Ansatz} further assumes that the same is true for $Q_z$. That is, our problem is to show, if possible, that \eqref{eq:RL12} implies  \eqref{eq:Ansatz}.

Let us assume for simplicity that $Q_z$ does not contain any elements of type (c),
\begin{equation}\label{eq:Qnoc}
Q_z \in  \sect{ \CC^{2}\angles{ L_1, L_2} \oplus \left(\liealg{u}(1)^3 \oplus \su(4)\right)^{\CC} }~.
\end{equation}
We may show that the $L_1$ and $L_2$ commute with $ \liealg{u}(1)^3 \oplus \liealg{su}(4)$. 
We may further assert that the preimage of $\left(\liealg{u}(1)^3 \oplus \su(4)\right)^{\CC}$ under the Lie bracket does not contain any $L_1\wedge (-)$ or $L_2\wedge (-)$ components. If we then restrict $Q_z$ to  $\left(\liealg{u}(1)^3 \oplus \su(4)\right)^{\CC}$ ,
\begin{equation}
\widetilde{Q}_z = \left. Q_{z} \right|_{ \left(\liealg{u}(1)^3 \oplus \su(4)\right)^{\CC}}~,
\end{equation}
we find that the curvature $R^{\widetilde{Q}}_{xy}$ of $\widetilde{Q}_z$,
\begin{equation}
R_{xy}^{\widetilde{Q}} = \partial_{x} \widetilde{Q}_{y}- \partial_{y} \widetilde{Q}_{x} + [\widetilde{Q}_x,\widetilde{Q}_y]~,
\end{equation} is identically zero,  $R^{\widetilde{Q}}_{xy}=0$, due to \eqref{eq:RL12} and the fact that the $L_1$, $L_2$ components in $Q_z$ are not relevant to 
\begin{equation}
R_{xy}^{\widetilde{Q}} \equiv  \left.  R_{xy}^{{Q}}\right|_{ \left(\liealg{u}(1)^3 \oplus \su(4)\right)}{} =0~.
\end{equation} 
If the space is simply connected, or we restrict to a simply-connected patch, there is a local $\liealg{u}(1)^3 \oplus \liealg{su}(4)$ gauge transformation that brings $\widetilde{Q}_z = 0$. Since this gauge transformation leaves $L_1$ and $L_2$ invariant, we are led to the Ansatz
\begin{equation}
Q_z = q_1 \, L_1 + q_2 \, L_2
\end{equation}
again, as a solution to \eqref{eq:dp2}, \eqref{eq:integrability4} and \eqref{eq:Qnoc}. The assumption \eqref{eq:Ansatz} essentially assumes that there are no elements of type (c) in $Q_z$ and any way to relax the assumption necessarily involves including such elements in $Q_z$. We leave it as an open question whether this is possible.


\section{\sfrac{3}{16}-BPS Solutions}
\label{sec:3bps}
In this section we show that if we assume $3$ complex Killing spinors that preserve 6 real supersymmetry out of 32, then the solution necessarily preserves a quarter of supersymmetry as in section \S\ref{sec:qbps}. The element $\Pz^0$ that was defined in \eqref{eq:reduction1} is a complex odd-degree form in the exterior algebra of
\begin{equation}
W = \RR^{5}\angles{ \tilde{e}_{4} ,   \tilde{e}_{5} ,\tilde{e}_{6} ,  \tilde{e}_{7} , \tilde{e}_8} \subset{U}~.
\end{equation}
The complex chiral spin representation of $\Spin(10)$ is the complex representation $\Delta_{10}\cong\Lambda^{\text{odd}}W^{\CC}\cong\CC^{16}$ and 
$\Pz^0$  is a complex chiral spinor of $\Spin(10)$. Note that in signature $(10,0)$, one cannot impose a reality condition on chiral spinors, only Majoranna and (complex) Weyl spinors exist similarly to four-dimensional Minkowski space.

The classification of spinors in $\Delta_{10}$ under the action of  $\Spin(10)$ is described in \cite{bryant}. 
There is a locally real two-dimensional orbit space $\mathcal{O}$ that is parametrized by a quadratic and a quartic on $\Delta_{10}$. A spinor $\Pz^0$ that is a representative of a generic orbit in $\mathcal{O}$ has stability subgroup $\SU(4)$ in $\Spin(10)$. At critical points of $\mathcal{O}$, the stability is enhanced to either $\SU(5)$ or $\Spin(7)$. In order to find the fixed form of a generic $\Pz^0$ with stability $\SU(4)$ we choose $\tilde{e}_4 \in \Lambda^{\text{odd}}W^{\CC}$ whose stability $\SU(4)\subset \Spin(10)$ is the same group we described in \eqref{eq:su4}. That is, $\SU(4)$ acts only on $\tilde{e}_5, \ldots \tilde{e}_8$. We may then decompose 
\begin{equation}
 \Lambda^{\text{odd}}W^{\CC} \stackrel{\SU(4)}{=} \CC\angles{\tilde{e}_4} \oplus \CC\angles{\tilde{e}_4\wedge\tilde{e}_5\wedge\tilde{e}_6\wedge\tilde{e}_7\wedge\tilde{e}_8} \oplus \CC^4 \oplus_{\pm} 2\Lambda^2_{\pm} \CC^4~.
\end{equation}
The $\CC^4$ are complex one-forms with no component along $\tilde{e}_4$ and the two $\Lambda^2_{\pm} W^{\CC}$ are complex three-forms with either one or no components along $\tilde{e}_4$, and with (anti-)self-duality with respect to $\SU(4)$ suitably imposed. We then identify the $\SU(4)$-invariant subspace 
in which $P_{z}$ belongs. This way, we assert that the generic ${P}_z^0$ is fixed to be
\begin{equation}\label{eq:spinor10}
\Pz^0 = a \, \tilde{e}_4 + b\, \tilde{e}_4\wedge\tilde{e}_5\wedge\tilde{e}_6\wedge\tilde{e}_7\wedge\tilde{e}_8~.
\end{equation}
The constants $a,b$ can be fixed to be real by using, for instance, the $U(1)\times U(1)$ that is generated by $e_4 \wedge e_{12}$ and $e_5 \wedge e_{13}$, see \eqref{eq:symsu} in the appendix. Let us summarize that the most general spinor $P_z^0$ of $\Spin(10)$ can be fixed to be of the form \eqref{eq:spinor10}.

Since the construction of the element \eqref{eq:spinor10} is not in \cite{bryant} in this form, let us elaborate on the enhancement of the stabilizer. If $a=0$ or $b=0$, then the stability is enhanced to $\SU(5)\subset \Spin(10)$. In the first case, $a=0$, we simply define $\SU(5)\subset\Spin(10)$ analogously to what we did for $\SU(4)$ in \eqref{eq:su4}, and which clearly preserves the volume form $\tilde{e}_4\wedge\tilde{e}_5 \wedge\tilde{e}_6\wedge\tilde{e}_7\wedge \tilde{e}_8$ and hence the special $\Pz^0$. In the second case, $b=0$, we may add to the generators of $\SU(4)$ the elements
\begin{align}
& e_4 \wedge e_{i+8} - e_i \wedge e_{12}~,\\
&e_4 \wedge e_i - e_{12} \wedge e_{i+8}~, \\
&e_4 \wedge e_{12} + e_5 \wedge e_{13}~,
\end{align}
for $i=5,6,7,8$. It is easy to show that they annihilate $\tilde{e}_4$, see \eqref{eq:spinhelp} for how these act, and so they also annihilate the special $\Pz^0$. If $|a|=|b|$, then the stability is enhanced to $\Spin(7)^+\subset\Spin(8)\subset\Spin(10)$, where $\Spin(7)^+$ leaves invariant $1+\tilde{e}_5 \wedge \ldots \tilde{e}_8$ as we described in section \S\ref{sec:qbps}.

In any case, it is clear that if $\Pz$ admits three complex supersymmetries. that is
\begin{equation}
- i\, e_{i} \cdot e_{i+8}  \cdot \Pz = \Pz ~, \quad i=1,2,3~,
\end{equation}
then it can be brought to the form
\begin{equation}
\Pz = \tilde{e}_1 \wedge \tilde{e}_2 \wedge \tilde{e}_3 \wedge \left(
 a \, \tilde{e}_4 + b\, \tilde{e}_4\wedge\tilde{e}_5\wedge\tilde{e}_6\wedge\tilde{e}_7\wedge\tilde{e}_8 \right)~.
\end{equation}
But then it also satisfies the projection condition
\begin{equation}
-i\, e_4 \cdot e_{12} \cdot \Pz = \left(-1+2 \tilde{e}_4 \wedge i_{\tilde{e}_4}\right)\Pz = \Pz~.
\end{equation}
Since the projection conditions are equivalent to the dilatino Killing spinor equation, which is in turn sufficient for supersymmetry because the gravitino Killing spinor is integrable on-shell, we conclude that $\sfrac{3}{16}$-susy solutions preserve at least four complex supersymmetries.

\section{Discussion}
\label{sec:discussion}
In this note we showed how the solvable Liouville field equations arise naturally for quarter-BPS solutions.  We consider it surprising that these suprsymmetric solutions comprise such a large class and are novel, although our derivation is relatively straightforward. Let us stress that we indeed used simple building blocks for our derivation: that a timelike background is ultrastatic \eqref{eq:ultrastatic1}, that one can easily construct projection operators on $P_z$ \eqref{eq:bps2}, and well-known facts about $\Spin(8)$ spinors in order to fix the projected $P_z$ completely \eqref{eq:fixing1}. Having reduced $P_z$ this way, the most obvious Ansatz \eqref{eq:Ansatz} for $Q_z$ produces uniquely an interesting class of solutions in terms of meromorphic functions \eqref{eq:sol1}-\eqref{eq:sol3}. The singularities of these functions then describe compatible quarter-BPS states. We have thus obtained a large class of solutions whose properties are amenable for an interesting analysis. It follows that the Ansatz itself is an interesting component to our analysis. A future direction is to try and see whether the Ansatz can be generalized or relaxed, a point on which we commented in section \S\ref{sec:ansatz}. 

We also showed that $\sfrac{3}{16}$-BPS solutions are necessarily $\sfrac{1}{4}$-BPS. This reproduces a result that stems from the classification of \cite{de_boer_classifying_2014}, that the timelike supersymmetric solutions can only preserve a fraction of $1$, $1/2$, $1/4$, $1/8$, or $1/16$ supersymmetries. In our case, setting $a$ or $b$ to zero enhances the supersymmetry from $1/4$ to $1/2$. Setting $a=b$, upon which the real stabilizer $\su(4)$ enhances to $\spin(7)$, should also be interesting but was not studied further. 

It is natural to look for generalizations of our solutions. There are two directions one may take. The first is to try to relax the Ansatz if possible, or else and presumably less likely to show that our solutions are in fact the unique quarter-BPS solutions if the Ansatz cannot be relaxed. The second direction involves studying solutions that preserve less than $\sfrac{3}{16}$-BPS solutions. The difficulty of the latter direction is that we know less about the $\Spin(16- 2\times 2)$-orbits of spinors, which is required if we are to repeat our method. Ideally, one would like to characterize all timelike supersymmetric solutions, i.e. preserving at least one timelike supersymmetry, which requires the classification of $\Spin(16-2\times 1)$ spinors. However, we have showed here that in principle a straightforward method may be fruitful as in the case of quarter-BPS solutions.

\subsection*{Acknowledgments}
The author acknowledges support from the Scientific and Technological Research Council of
Turkey (T\"UB\.ITAK) project 113F034.

\appendix
\section{Linearity of Timelike Killing Spinors}
\label{app:lineps}

Assume $n$ Killing spinors $\epsilon_z^i$, $i=1,\ldots n$, that are $\CC$-linearly independent. We may bring them to the form of \eqref{eq:LD} by $\SO(16)$ rotations. First we rotate $\epsilon^1_z$ so that the real and imaginary part span the directions $e_1$ and $e_9$. Because it is complex null, it has to be of the form in \eqref{eq:LD}: 
\begin{equation}
\epsilon_z^1=N_{11}\left(e_1+ i e_9\right)~.
\end{equation}
Then rotate $\epsilon_z^2$ using the stabilizer $\SO(16-2)$ of $\epsilon_z^1$ so  that the real and imaginary part span the directions $e_1$, $e_2$, $e_9$ and $e_{10}$. Because $\epsilon_z^2$ is orthogonal to $\epsilon_z^1$ and null itself, the coefficients of $\epsilon_z^2$ have to be related as in \eqref{eq:LD},
\begin{equation}
\epsilon_z^2=N_{21}\left(e_1+ i e_9\right) + N_{22}\left(e_2+ i e_{10}\right)~.
\end{equation} 
We proceed analogously for the rest, the $k$-th spinor has to be null along $e_{i}+i e_{i+8}$, $i\leq k$, and so 
\begin{equation}\label{eq:LD2}
\epsilon_z^k=\sum_{i=1}^{k}N_{ki}\left( e_{i}+i e_{i+8} \right)~.
\end{equation}
At this point, we have assumed that $N_{ii}\neq0$ in our iteration. 

We will prove now that in \eqref{eq:LD2}, the diagonal coefficient $N_{kk}$ cannot be zero. Assume there is a coefficient $N_{kk}=0$, for some $k>1$, and $N_{ii}\neq 0$ for all $i<k$. Then clearly the k-th spinor $\epsilon_z^i$ is $C(M)$-linearly dependent on the $(k-1)$ spinors $\epsilon_z^i$, $i<k$,
\begin{equation}
\epsilon_z^k = \sum_{i=1}^{k-1} c_i(z,\bar{z}) \epsilon_z^i~.
\end{equation}
But since the $\epsilon_z^i$, $i<k$, are $C(M)$-linearly independent, by taking the gravitino Killing spinor equations and using the Leibniz identity, we may show that the $c_i$ are constants. The $k$-th Killing spinor is thus $\CC$-linearly dependent on its predecessors. 


\section{Spinor Basis}
\label{app:spinhelp}


Elements of the spin algebra in the Clifford representation \eqref{eq:clifford1} appear often in the following combinations
\begin{subequations}\label{eq:spinhelp}
\begin{align}
e_i \wedge e_j + e_{i+8} \wedge e_{j+8} &= -8 \tilde{e}_{[i} \wedge i_{\tilde{e}_{j]}}  ~, \\
e_i \wedge e_j - e_{i+8} \wedge e_{j+8} &= 4\left( \tilde{e}_i \wedge \tilde{e}_j +  i_{\tilde{e}_i} i_{\tilde{e}_j}\right) ~,\\
e_i \wedge e_{j+8} + e_j \wedge e_{i+8} &= 4 i \left(- \delta_{ij} + 2 \tilde{e}_{(i}\wedge i_{\tilde{e}_{j)}} \right) ~, \label{eq:symsu}\\
e_i \wedge e_{j+8} - e_j \wedge e_{i+8} &= 4 i \left( \tilde{e}_i \wedge \tilde{e}_j -  i_{\tilde{e}_i} i_{\tilde{e}_j} \right)~.
\end{align}\end{subequations}
Note that in these equalities we use the natural image
\begin{equation}
a\wedge b = a \otimes b - b \otimes a 
 \stackrel{\Cl(V_{16})}{\longmapsto} a \cdot b  - b\cdot a ~.
\end{equation}
On the other hand, the spin representation is minus one quarter the Clifford action, see \eqref{eq:spinrep}. This one quarter is included on the right-hand side of \eqref{eq:ua}, \eqref{eq:su4} and \eqref{eq:ub}.

\providecommand{\href}[2]{#2}\begingroup\raggedright\endgroup

\end{document}